# Spatial filtering of audible sound with acoustic landscapes

Shuping Wang[1], Jiancheng Tao[1,*], Xiaojun Qiu[2], Jianchun Cheng[1]

[1] Key Laboratory of Modern Acoustics and Institute of Acoustics, Nanjing University, Nanjing, China

[2] Centre for Audio, Acoustics and Vibration, Faculty of Engineering and IT, University of Technology Sydney, Sydney, Australia

*Correspondence to: jctao@nju.edu.cn

Acoustic metasurfaces manipulate waves with specially designed structures and achieve properties that natural materials cannot offer. Similar surfaces work in audio frequency range as well and lead to marvelous acoustic phenomena that can be perceived by human ears. Being intrigued by the famous Maoshan Bugle phenomenon, we investigate large scale metasurfaces consisting of periodic steps of sizes comparable to the wavelength of audio frequency in both time and space domains. We propose a theoretical method to calculate the scattered sound field and find that periodic corrugated surfaces work as spatial filters and the frequency selective character can only be observed at the same side as the incident wave. Maoshan Bugle phenomenon can be well explained with the method. Finally, we demonstrate that the proposed method can be used to design acoustical landscapes, which transform impulsive sound into famous trumpet solos or other melodious sound.

Metasurfaces manipulate waves with specially designed structures and achieve properties that natural materials cannot offer.[1-6] Acoustic metasurfaces of subwavelength thickness provide many kinds of wave manipulation such as acoustic one-way transmission,[7-9] negative refraction of sound waves,[10] acoustic cloaking[11-13] and so on[14-19]. Periodic corrugated surfaces consisting of steps with sizes comparable to the wavelength of audible sound can be considered as equivalents of metasurfaces in audible frequency range, which lead to a lot of marvelous acoustic phenomena. For example, the steps in front of the Southern Jiangsu Victory Monument can produce 6 bugle-like notes as a result of a firecracker being set off.[20-21] Similar phenomena exist in the El Castillo pyramid, where a sound echo like the chip of a Quetzal bird is produced in response to a handclap and observers hear pulses that sound like raindrops falling in a water filled bucket when other people are climbing the pyramid higher up.[22]

Theories like Fresnel-Kirchhoff diffraction, Bragg reflection, leaky Rayleigh waves, acoustic negative reflection and finite element method have been used to investigate these phenomena;[20-25] however, none of the previous work reports accurate theoretical solution to the scattered sound field either in frequency or time domain. The mechanisms and physical images of these kinds of large scale metasurfaces are not clear and there is no approach or procedure reported in previous work to design such a landscape to generate specified sound.

In this letter, we propose an analytical method to calculate the sound field scattered by a periodic corrugated surface and obtain some physical insights into the problem. Numerical simulations and experiments are carried out to demonstrate its validity. We investigate the spatial distribution of the fundamental frequency of scattered sound and their relations to the step sizes. Finally, two acoustical landscapes are designed based on the proposed method which turn impulsive sound into famous trumpet solos and proves that audible sound can be manipulated by artificial surfaces. Our proposed method extends metasurfaces to audio frequency range and provides the possibility to design interesting acoustical landscapes which have promising applications.

Figure 1(a) is a schematic diagram of such an acoustical landscape which consists of $N$ groups of steps of different sizes. A series of $N$ notes can be heard at the position of the listener in response of an impulse at the position of the source. Figure 1(b) is a cross-sectional view of one of the groups of steps.

According to the Kirchhoff-Helmholtz equation, sound scattered by the surface can be expressed as an integral over the surface A (the surface of the steps)[26, 27]

$$p_s = \int_A \left[ G \frac{\partial p}{\partial \boldsymbol{n}} - p \frac{\partial G}{\partial \boldsymbol{n}} \right] dx, \qquad (1)$$

where $G$ is the Green's function in semi-infinite space and $\boldsymbol{n}$ is the normal vector on the surface. Once the total sound pressure $p$ on the surface is obtained, the scattered sound can be calculated (see Supplementary information for more detailed process).

We applied the theoretical method on two examples to demonstrate its feasibility. The first one is 10 steps of the size $h = 0.125$ m and $l = 0.315$ m. The theoretical scattered sound pressure level calculated in frequency domain is shown in Fig. 2(a). We can see that the steps work as a frequency-selective filter and passband is around 534 Hz and its harmonics. The wavelength of the fundamental frequency (340/534 = 0.64 m) is about twice the dimensions of the steps ($h$ = 0.125 m, $l$ = 0.315 m, $L$ = 0.339 m). The boundary element model of the steps was created in ANSYS 14.5 and imported into Sysnoise 5.6, a commercial boundary element software to obtain the simulation scattered sound pressure, and the results are also shown in Fig. 2(a), which agrees well with theoretical results at most frequencies below 3000 Hz.

We carried out experiments in the anechoic chamber of Nanjing University to further demonstrate simulation results. The experimental setup is indicated in Fig. 2(b). 10 steps of the same length and height as in the above simulations are constructed with boards, and the width is 3.0 m. In the experiments, the starting pistol was used to generate an impulse that is sufficiently short in time domain so that sound scattered by the steps can be separated from direct sound and that reflected from the baffle. A B&K PULSE 3560B analyzer was used to record the signal at the position of the receiver. The recorded signal in time domain is shown in Fig. 2(c). The signal in the green box is the direct and reflected sound and sound scattered by the steps is in the red box. The time delay between the direct and scattered sound is about 0.013 s which corresponds to the difference of the length of acoustic paths (about 4.5 m). The power spectra density of the scattered sound in the red box is also indicated in Fig. 2(c). The fundamental frequency of the experimental scattered sound is about 544 Hz and the error with that predicted by the proposed method (534 Hz) is 1.87%.

After transforming the theoretical scattered sound from frequency to time domain and convolving with the recorded impulse of the starting pistol, we obtain the theoretical time domain signal and it is shown in Fig. 2(c) as well as the power spectra density of the scattered sound. It is observable that the theoretical scattered signal agrees well with experimental results in both frequency and time domains.

Another example we investigate is the Mount Maoshan model (see Fig. S1 in the Supplementary information for the picture of Mount Maoshan monument site and the cross-sectional view of the steps). There are 6 groups of steps, and the steps in the same group have about the same sizes which are listed in Table S1 in the Supplementary information. Fundamental frequencies of the sound scattered by each of the 6 groups of steps obtained by applying the proposed method are listed in Table I. Numerical simulation method was also applied on this example by using Odeon 11.23, a room acoustics software and the results are listed in Table I, as well as the measured results. It is clear that the fundamental frequencies obtained with the proposed method, Odeon and measurements agree well with each other, and the maximum error is 4.35%. We also compared the theoretical and recorded scattered sound signal in time domain, which sound similarly and proves the validity of the proposed method.

The acoustic characters of periodic steps are further investigated by using the proposed method and two interesting phenomena are observed. First, the maximum scattered sound pressure level at a certain frequency appears within a certain direction at the same side as the incident wave. Figures 3(a) shows the incident sound field when the point source is located at (8.89, 2.00) m, which is marked as a star. Figures 3(b)-3(d) show the distribution of scattered sound pressure level within the area −6 m $< x <$ 10 m, 0 m $< y <$ 3 m at different exciting frequencies (500 Hz, 600 Hz and 700 Hz). 10 steps are located from $x = 0$ m to $x = 3.39$ m as indicated in Fig. 1(b). The length of each step is 0.315 m and the height is 0.125 m. In Figs. 3(b)-3(d), the area where maximum sound pressure level appears is near the space indicated by the arrow, which is at different directions at different exciting frequencies; besides, this area is always at the same side as the incident wave.

Secondly, the fundamental frequencies of scattered sound field have a spatial distribution. Figure 4(a) shows the fundamental frequencies at different locations when the sound source is fixed at (63.16, 70) m and 10 steps of size $h = 0.1$ m, $l = 0.3$ m are located from $x = 0$ m to $x = 3.16$ m. We can see that the fundamental frequency increases with $x$ ($d_r$) and decreases with $y$ ($h_r$) which indicates a spatial filtering mechanism here. Of course the fundamental frequency of sound scattered by periodic steps is also related to the size of the steps ($h$ and $l$). Figure 4(b) shows the fundamental frequencies of scattered sound when the dimensions of the steps ($h$ and $l$) change while the relative positions of the sound source and receiver to the 10 steps are fixed ($d_s = 60$ m, $h_s = 70$ m, $d_r = 30$ m, $h_r = 35$ m). We can find that the fundamental frequency decreases with either $h$ or $l$.

Besides predicting the sound field scattered by periodic corrugated steps, we can apply the proposed method to design acoustical landscapes. Theoretically, every kind of sound can be generated at a certain location by careful design of the metasurfaces and exciting signal. The detailed procedure is as follows:

(1) Analyze the desired sound to be generated by the surfaces and obtain the

fundamental frequencies. If there are *N* notes, there should be *N* groups of steps.

(2) Choose positions of the sound source and listener. As mentioned above, periodic steps work as spatial filters and the fundamental frequencies of scattered sound vary at different locations. The sound source and listener must be at the same side of the steps.

(3) Start from the step group nearest to the sound source and listener. Estimate the initial size of the steps (*h* and *l*) with

$$l = \frac{c_0}{2f_0}, \tag{2}$$

$$h = \frac{l}{3}, \tag{3}$$

where $c_0$ is the sound speed and $f_0$ is the desired fundamental frequency.

(4) Calculate the response in frequency domain using the proposed method and then transfer it to time domain by inverse Fourier transformation, or directly simulate the impulse response using commercial acoustic softwares such as Odeon.

(5) Calculate the scattered sound in time domain by convolving the impulse response with an exciting signal and adjust the parameters *h* and *l* with listening tests until the note sounds satisfying.

(6) Repeat the process for the rest of the groups using similar procedure as that for the first group. The distance (Δ*L*) from the current to the previous step group needs to be estimated first with the time interval *t* between two notes by Δ*L* = $c_0 t$/2. This will determine the distance from the group to the sound source ($d_s$). With this initial value, we can apply the same procedure as mentioned above in Steps 3-5 to obtain the parameters of each group. The parameters Δ*L*, *h* and *l* need to be adjusted simultaneously with listening tests until all the notes and the time intervals between them sound satisfying.

Two trumpet solos are designed as examples: "Toreador song" and "Neapolitan dance". They consist of 10 notes and 8 notes, respectively, and the fundamental frequencies of the notes are listed in Tables S2 and S3 in the supplementary information. The proposed method is used to design the size of each step group and the distance between them (The model of the steps is indicated in Fig. 1(a)). The optimized parameters of the steps are listed in Tables S2 and S3 as well. The source is 60 m and the listener is 58 m away from the nearest step group in $x$ direction. The height of the source and listener is 70 m and 35 m, respectively.

In summary, we propose a theoretical method which can be used to calculate the sound field scattered by a structure consisting of periodic corrugated steps. The feasibility of the method is demonstrated by numerical simulations and experiments. The model is two-dimensional, but it can be used to predict fundamental frequencies of the sound scattered by three-dimensional steps with much less computation load (The error between fundamental frequencies obtained from two-dimensional and three dimensional model is 0.37%, see supplementary information). Although similar phenomena exist in nature, it is the first time that a theoretical model is proposed to describe it in both space and time domains. By using the proposed method, we find that periodic steps work as spatial filters and the scattered sound is different at different positions with the same exciting signal. We also propose a procedure to design interesting acoustical landscapes which is ready for real applications. With comparable size to the wavelength of audio frequencies, such a structure leads to marvelous acoustic phenomena which can be perceived by human ears.

**Acknowledgments**

This work was supported by the National Natural Science Foundation of China (No. 11474163 and 11634006). The authors would also like to thank Mr. Kang Wang for his help with Odeon simulations.

TABLE I. The fundamental frequency of the sound scattered by the 6 groups of steps obtained by the proposed method, Odeon simulation and measurements.

| The fundamental frequency (Hz) | Group number | | | | | |
|:---:|:---:|:---:|:---:|:---:|:---:|:---:|
| | 1 | 2 | 3 | 4 | 5 | 6 |
| The proposed method | 407 | 380 | 368 | 568 | 558 | 427 |
| Odeon simulation | 399 | 368 | 363 | 565 | 554 | 424 |
| Measurements | 416 | 384 | 368 | 560 | 552 | 424 |

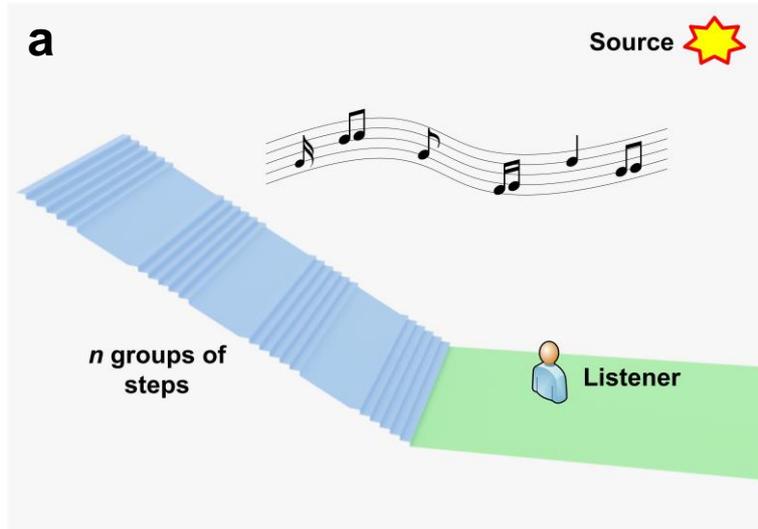

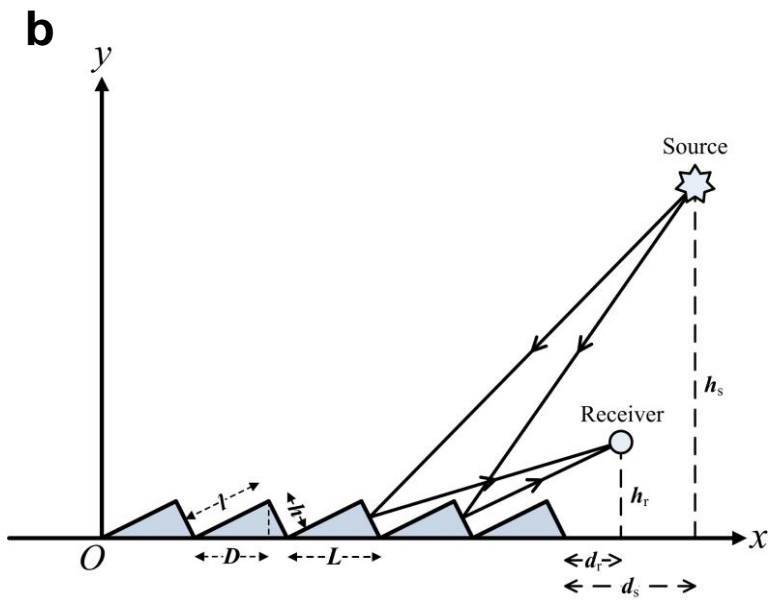

FIG. 1. (a) Schematic diagram of an acoustical landscape consisting of steps. (b) Cross-sectional view of a group of periodic steps. The length and height of each step is $l$ and $h$, respectively, and the surfaces of the steps are rigid. The distance between the steps and source is $d_s$ along $x$ axis, and the distance between the steps and receiver is $d_r$ along $x$ axis. The height of the source and receiver is $h_s$ and $h_r$, respectively.

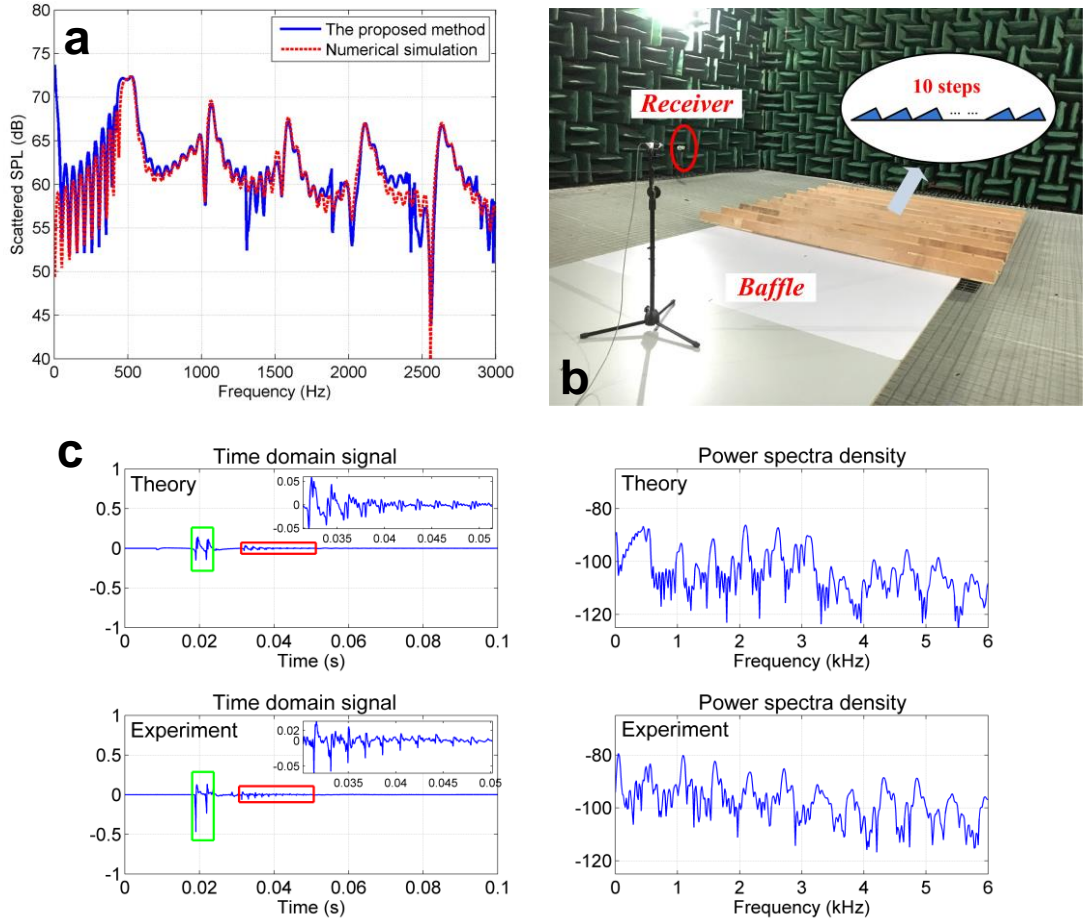

FIG. 2. (a) The scattered sound pressure level (SPL) obtained by the proposed theoretical method and numerical simulations. $d_s$ = 5.5 m, $h_s$ = 2.0 m, $d_r$ = 2.0 m, $h_r$ = 1.0 m. (b) The experimental setup in the anechoic chamber. (c) The theoretical and experimental scattered signals in time domain and its power spectra density.

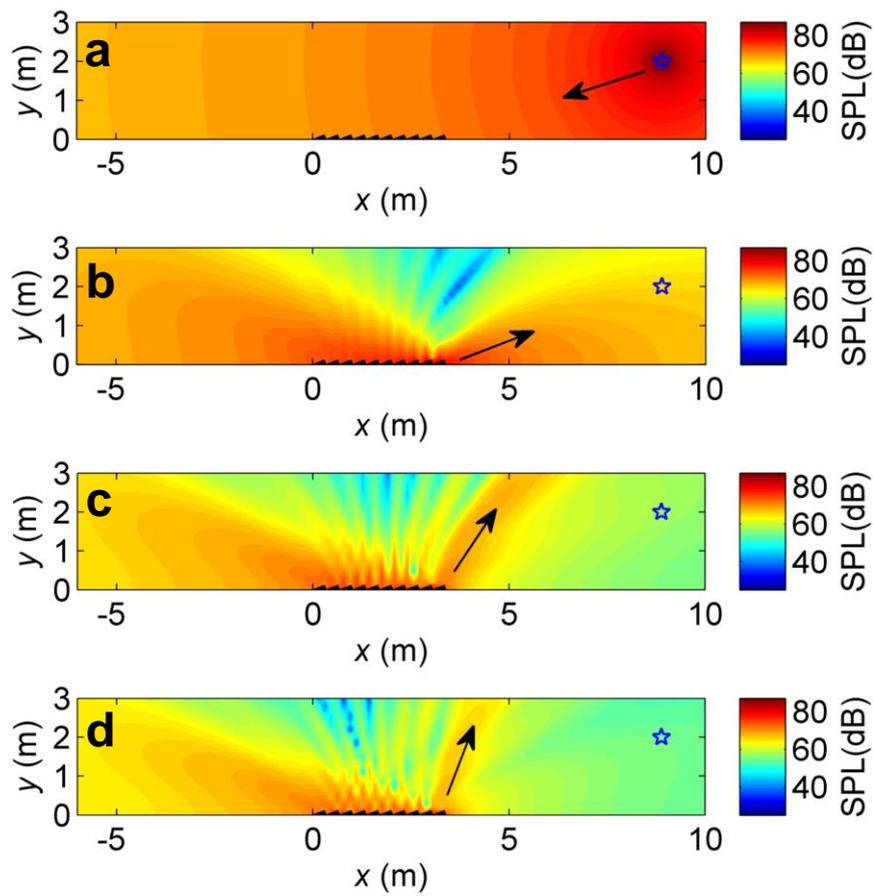

FIG. 3. The distribution of the incident and scattered sound pressure level within the area −6 m < x < 10 m, 0 m < y < 3 m where the sound source is marked as a star. (a) Incident sound. (b) Scattered sound at 500 Hz. (c) Scattered sound at 600 Hz. (d) Scattered sound at 700 Hz.

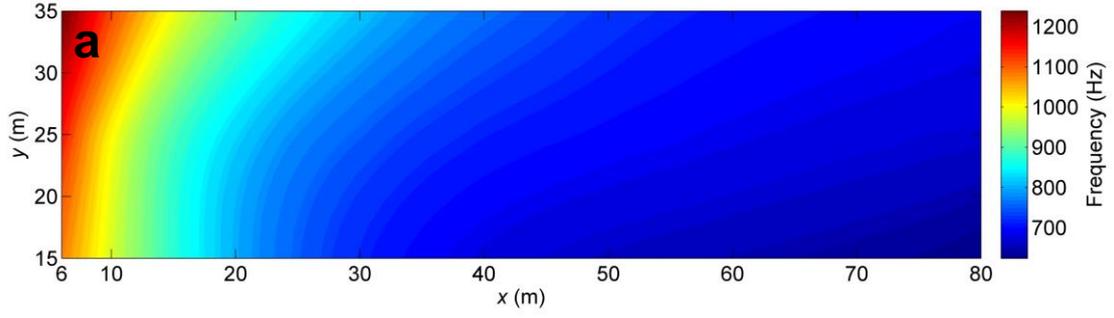

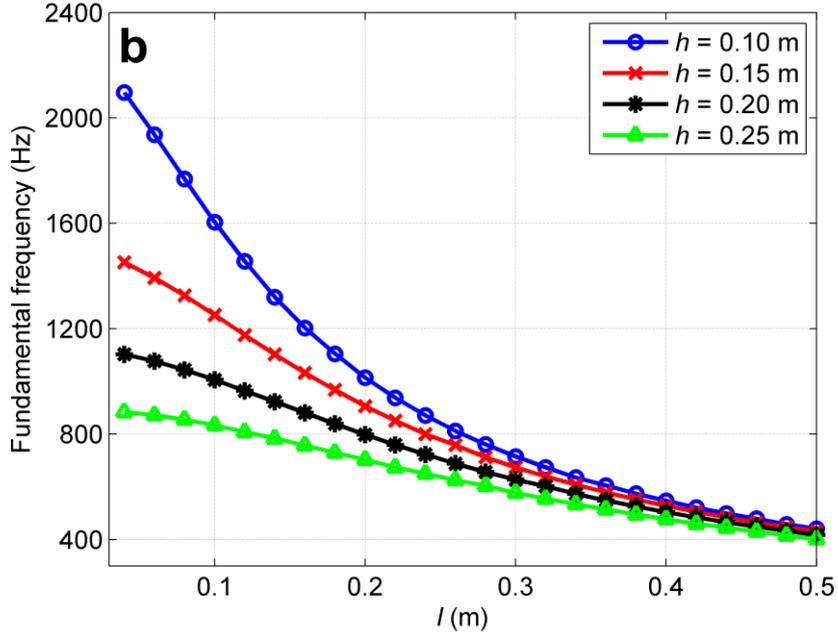

FIG. 4. (a) The spatial distribution of fundamental frequency within the area 6 m < $x$ < 80 m, 15 m < $y$ < 35 m when the sound source is fixed at (63.16, 70) m and 10 steps of size $h$ = 0.1 m, $l$ = 0.3 m are located from $x$ = 0 m to $x$ = 3.16 m. (b) The fundamental frequencies of the scattered sound pressure when $h$ = 0.1 m, $l$ = 0.3 m, $d_s$ = 60 m, $h_s$ = 70 m.

**Supplementary information**

**Detailed process of the proposed method**

In Fig. 1(b), the height of the $n$th step is

$$\xi(x) = \begin{cases} \dfrac{h}{D}x, & nL < x \leq nL+D \\ -\dfrac{h}{D}(x-nL), & nL+D < x \leq nL+L \end{cases}, \tag{S1}$$

where

$$L = \sqrt{h^2 + l^2}, \tag{S2}$$

$$D = \frac{l^2}{L}. \tag{S3}$$

The incident wave at location $r$ is the sound field radiated by an infinitely long vibrating cylinder[26]

$$p_i = AH_0^{(2)}(k|\bm{r}-\bm{r}_s|), \tag{S4}$$

where $A$ is the source strength, $H_0^{(2)}$ is the Hankel Function of the second type, $k$ is the wave number and $r_s$ is the position of the source.

According to the image method, the reflected sound field at location $r$ is

$$p_r = AH_0^{(2)}(k|\bm{r}-\bm{r}_s'|), \tag{S5}$$

where $r_s$' is the image location of $r_s$ regarding the rigid plane $y = 0$.

The scattered sound field is a line integral over area A (the surface of the steps) and $\Sigma$ (the surface of the baffle)[27]

$$p_s = \int_{A+\Sigma}\left[G\frac{\partial p}{\partial \bm{n}} - p\frac{\partial G}{\partial \bm{n}}\right]dx, \tag{S6}$$

in which $\bm{n}$ is the normal vector on the surface and $G$ is the Green's function for a rigid baffle at $y = 0$

$$G = -\frac{j}{4}H_0^{(2)}(kr) - \frac{j}{4}H_0^{(2)}(kr'), \tag{S7}$$

where $r$ is the distance between the receiver and the source and $r'$ is the distance between the receiver and the image source.

Because the surfaces of the steps and baffle are all rigid, Eq. (S6) can be simplified as

$$p_s = \int_A -p\frac{\partial G}{\partial \boldsymbol{n}}dx, \tag{S8}$$

and this is an integral over area A. It is difficult to obtain the analytical solution of Eq. (S8), but we can get the accurate solution after discretization operation if the discretization interval is small enough.

Equation (S8) is further written as an integral on A', the projection of A on $x$-axis

$$p_s = \int_{A'}\left[-p\left(\frac{\partial \xi}{\partial x}\frac{\partial G}{\partial x} - \frac{\partial G}{\partial y}\right)\right]dx. \tag{S9}$$

At the same time, $p_s = p - p_i - p_r$, thus the total sound pressure at location $\boldsymbol{r}$ can be obtained by solving the following integral equation

$$p - p_i - p_r = \int_{A'}\left[-p\left(\frac{\partial \xi}{\partial x}\frac{\partial G}{\partial x} - \frac{\partial G}{\partial y}\right)\right]dx. \tag{S10}$$

It is difficult to derive a closed form solution of Eq. (S10), therefore numerical scheme is utilized, and the integral equation is discretized as

$$\frac{1}{2}p(x_i) - p_i(x_i) - p_r(x_i) = \sum_{j=1}^{M} -p(x_j)\left[\frac{\partial \xi}{\partial x}(x_j)\frac{\partial G}{\partial x}(x_i|x_j) - \frac{\partial G}{\partial y}(x_i|x_j)\right]\Delta x_j \tag{S11}$$

on the surface of the steps, and it can be described in the form of matrix as

$$\frac{1}{2}\boldsymbol{P} - \boldsymbol{P}_i - \boldsymbol{P}_r = \boldsymbol{AP}, \tag{S12}$$

where $\boldsymbol{P} = [p(x_1, \xi(x_1)), p(x_2, \xi(x_2)), \ldots p(x_M, \xi(x_M))]^T$, $\boldsymbol{P}_i = [p_i(x_1, \xi(x_1)), p_i(x_2, \xi(x_2)),$

...$p_i(x_M, \xi(x_M))]^T$, $\boldsymbol{P}_r = [p_r(x_1, \xi(x_1)), p_r(x_2, \xi(x_2)), ...p_r(x_M, \xi(x_M))]^T$,

$$A = -\Delta x \begin{pmatrix} \frac{\partial \xi}{\partial x}(x_1)\frac{\partial G}{\partial x}(x_1|x_1) - \frac{\partial G}{\partial y}(x_1|x_1) & \frac{\partial \xi}{\partial x}(x_2)\frac{\partial G}{\partial x}(x_1|x_2) - \frac{\partial G}{\partial y}(x_1|x_2) & \cdots & \frac{\partial \xi}{\partial x}(x_M)\frac{\partial G}{\partial x}(x_1|x_M) - \frac{\partial G}{\partial y}(x_1|x_M) \\ \frac{\partial \xi}{\partial x}(x_1)\frac{\partial G}{\partial x}(x_2|x_1) - \frac{\partial G}{\partial y}(x_2|x_1) & \frac{\partial \xi}{\partial x}(x_2)\frac{\partial G}{\partial x}(x_2|x_2) - \frac{\partial G}{\partial y}(x_2|x_2) & \cdots & \frac{\partial \xi}{\partial x}(x_M)\frac{\partial G}{\partial x}(x_2|x_M) - \frac{\partial G}{\partial y}(x_2|x_M) \\ \vdots & \vdots & \ddots & \vdots \\ \frac{\partial \xi}{\partial x}(x_1)\frac{\partial G}{\partial x}(x_M|x_1) - \frac{\partial G}{\partial y}(x_M|x_1) & \frac{\partial \xi}{\partial x}(x_2)\frac{\partial G}{\partial x}(x_M|x_2) - \frac{\partial G}{\partial y}(x_M|x_2) & \cdots & \frac{\partial \xi}{\partial x}(x_M)\frac{\partial G}{\partial x}(x_M|x_M) - \frac{\partial G}{\partial y}(x_M|x_M) \end{pmatrix}. \quad (S13)$$

The total sound pressure $p(x_j)$ at $(x_j, 0)$ on the surface of the steps is

$$\boldsymbol{P} = 2(\boldsymbol{I} - 2\boldsymbol{A})^{-1}(\boldsymbol{P}_i + \boldsymbol{P}_r), \quad (S14)$$

and the scattered sound pressure at any location $\boldsymbol{r} = (x, y)$ is

$$p_s(x, y) = \sum_{j=1}^{M} -p(x_j)\left[\frac{\partial \xi}{\partial x}(x_j)\frac{\partial G}{\partial x}(x, y|x_j) - \frac{\partial G}{\partial y}(x, y|x_j)\right]\Delta x. \quad (S15)$$

By applying IFFT on Eq. (S15) and convolving with the exciting signal, the scattered sound in time domain can be obtained.

**Comparison between two dimensional and three dimensional models**

Two dimensional (2D) and three dimensional (3D) models are compared and the results are shown in Fig. S2. These two models consist of 10 steps of the same size $h$ = 0.125 m, $l$ = 0.315 m. $d_s$ = 5.5 m, $h_s$ = 2.0 m, $d_r$ = 2.0 m, $h_r$ = 1.0 m. In the three dimensional model, the width of the steps is 3.0 m and the source and listener are at the position of 1.5 m, as indicated in Fig. S2a. Figure S2b shows comparison between the scattered sound pressure level of a two dimensional (2D) model predicted by the proposed method and a three dimensional (3D) model of the same size obtained by numerical simulations using Sysnoise 5.6.

We can see from Fig. S2b that the fundamental frequency of the three dimensional and two dimensional model is 535 Hz and 533 Hz, respectively, which matches well, and the error is only 0.37%. So we can use the two dimensional model to predict the fundamental frequencies of the scattered sound of the real three dimensional model, which is accurate enough and at the same time, reduces the computation load.

**Supplementary Figures**

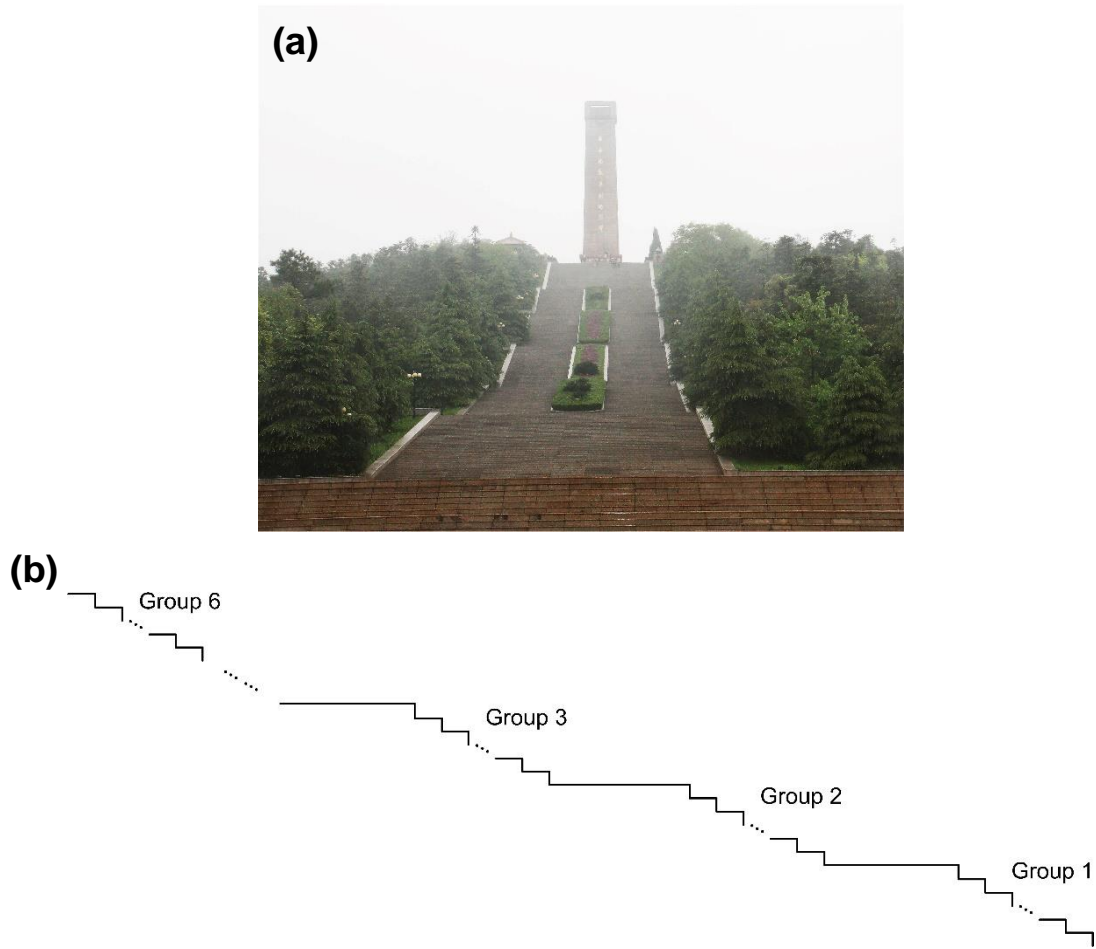

FIG. S1. (a) Picture of the steps in Mount Maoshan, where the structure in front of the monument consists of 6 groups of steps of different sizes. (b) Cross-sectional view of the 6 groups of steps.

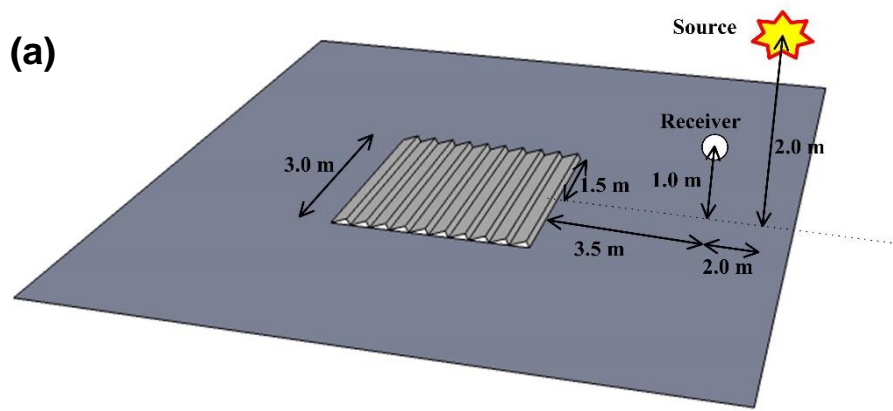

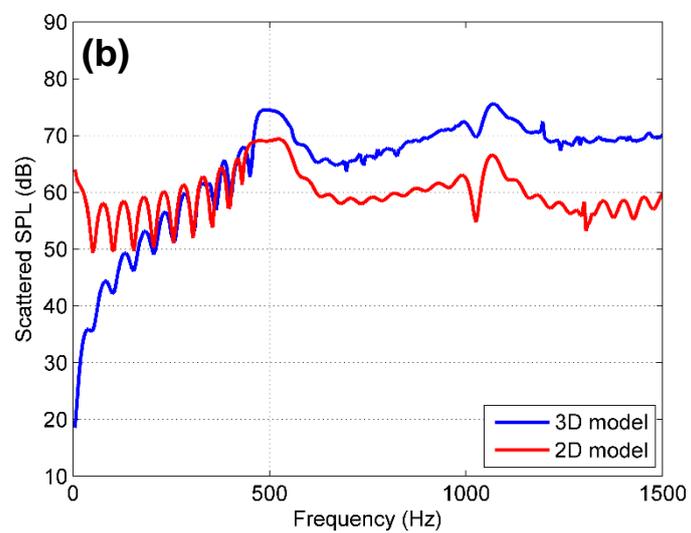

FIG. S2. (a) The schematic of the three dimensional model. (b) The scattered sound pressure level (SPL) in the three dimensional and two dimensional model.

**Supplementary Tables**

TABLE S1 The parameters of the steps in Mount Maoshan.

| Group number | Number of steps | Height ($\Delta h$)/m | Length ($\Delta l$)/m | Length of the platform between adjacent step groups ($L$)/m |
|---|---|---|---|---|
| 1 | 49 | 0.10 | 0.52 | 10.12 |
| 2 | 50 | 0.10 | 0.49 | 10.09 |
| 3 | 50 | 0.14 | 0.48 | 10.68 |
| 4 | 50 | 0.15 | 0.31 | 7.51 |
| 5 | 49 | 0.13 | 0.31 | 9.91 |
| 6 | 60 | 0.15 | 0.40 | N/A |

TABLE S2 The parameters of the 10 groups of steps and the fundamental frequencies corresponding to each step group for the design of "Toreador song".

| Group number | Number of steps | Height ($\Delta h$)/m | Length ($\Delta l$)/m | Length of the platform between adjacent step groups ($\Delta L$)/m | Fundamental frequency/Hz |
| --- | --- | --- | --- | --- | --- |
| 1 | 35 | 0.100 | 0.271 | 41.666 | 784 |
| 2 | 45 | 0.075 | 0.199 | 43.445 | 880 |
| 3 | 30 | 0.075 | 0.213 | 18.112 | 784 |
| 4 | 30 | 0.090 | 0.250 | 65.888 | 659 |
| 5 | 35 | 0.080 | 0.240 | 60.013 | 659 |
| 6 | 35 | 0.080 | 0.240 | 70.112 | 659 |
| 7 | 40 | 0.090 | 0.260 | 3.768 | 587 |
| 8 | 40 | 0.085 | 0.220 | 47.267 | 659 |
| 9 | 55 | 0.075 | 0.220 | 8.256 | 698 |
| 10 | 60 | 0.080 | 0.232 | N/A | 659 |

TABLE S3 The parameters of the 8 groups of steps and the fundamental frequencies corresponding to each step group for the design of "Neapolitan dance".

| Group number | Number of steps | Height ($\Delta h$)/m | Length ($\Delta l$)/m | Length of the platform between adjacent step groups ($\Delta L$)/m | Fundamental frequency/Hz |
| --- | --- | --- | --- | --- | --- |
| 1 | 20 | 0.190 | 0.540 | 48.326 | 392 |
| 2 | 15 | 0.140 | 0.413 | 30.458 | 440 |
| 3 | 15 | 0.120 | 0.325 | 24.733 | 523 |
| 4 | 20 | 0.115 | 0.340 | 26.897 | 494 |
| 5 | 35 | 0.130 | 0.360 | 13.104 | 440 |
| 6 | 40 | 0.140 | 0.410 | 5.670 | 392 |
| 7 | 80 | 0.115 | 0.312 | 130.724 | 494 |
| 8 | 150 | 0.110 | 0.311 | N/A | 494 |